\documentclass[twocolumn,epjc3]{svjour3}
\smartqed  
\usepackage{amsmath, empheq}
\usepackage{amsfonts}
\usepackage{amssymb}
\usepackage{marvosym}
\usepackage{breqn}
\usepackage{wasysym}
\usepackage[font=small,labelfont=bf]{caption}
\usepackage{subcaption}
\usepackage{fullpage}
\usepackage[]{graphicx}
\usepackage{hyperref}
\usepackage[english]{babel}
\usepackage{xcolor}
\usepackage{makeidx}
\numberwithin{equation}{section}

\newcommand{\be}{\begin{eqnarray}}
\newcommand{\ee}{\end{eqnarray}}

\hypersetup{citebordercolor=0 1 0, linkbordercolor=1 1 1,}

\journalname{Eur. Phys. J. C}

\begin{document}
\title{Superheavy dark matter in  $R+R^2$ cosmology with conformal anomaly
}

\author{E. V. Arbuzova\thanksref{e1,addr1,addr2}
\and
A. D. Dolgov\thanksref{e2,addr2,addr3}
\and
R. S. Singh\thanksref{e3,addr2}
}

\thankstext{e1}{e-mail: arbuzova@uni-dubna.ru}
\thankstext{e2}{e-mail: dolgov@fe.infn.it}
\thankstext{e3}{e-mail: akshalvat01@gmail.com}

\institute{Department of Higher Mathematics, Dubna State University, Universitetskaya ul. 19, Dubna 141983, Russia \label{addr1} \and Department of Physics, Novosibirsk State University, Pirogova 2, Novosibirsk 630090, Russia \label{addr2} \and ITEP, Bol. Cheremushkinskaya 25, Moscow 117218, Russia \label{addr3} }

\date{Received: date / Accepted: date}

\maketitle

\begin{abstract}
Cosmological evolution and particle creation in $R^2$-modified gravity are considered for the case of the dominant 
decay of the scalaron into a pair of gauge bosons due to conformal anomaly. It is shown that in the process 
of thermalization superheavy dark matter with the coupling strength typical for the GUT SUSY can be created.
Such dark matter would have the proper cosmological density if the particle mass is close to $10^{12}$ GeV.
\end{abstract}

\pagenumbering{arabic}

\section{Introduction \label{s-Intro}}

The most popular and natural hypothesis that dark matter consists of the lightest supersymmetric particles {(LSP)}
somewhat lost its popularity since no manifestation of supersymmetry (SUSY) was observed  at LHC~\cite{LHC}. 
The LHC data significantly restricted parameter space open for SUSY. Though strictly speaking low energy SUSY, around 
1 TeV, is not excluded and no direct limits from below on the LSP mass were presented, see \cite{pdg}, still a study
of higher energy SUSY and heavier LSPs can be of interest.

Different mechanisms of LSP production in cosmology are summarized in Ref.~\cite{SUSY-rev}.
If they behave as the the usual WIMPs, then their frozen number density is governed by the Zeldovich 
equation~\cite{Zeldovic:1965}, see also~\cite{LW,VDZ}, and their energy density 
according to the conventional calculations, is
\be
\rho_{LSP} \sim \rho_{DM}^{(obs)} (M_{LSP}/ 1\,\rm{TeV})^2 ,
\label{rho-LSP}
\ee
where  $ M_{LSP} $ is the mass of LSP and $\rho_{DM}^{(obs)} \approx 1$ keV/cm$^3$ is the observed 
value of the cosmological density of dark matter.

This equation is based on the general result first derived by Zeldovich~\cite{Zeldovic:1965} and later 
rederived in detail in several textbook, see e.g. ~\cite{Kolb,GR-v1}:
\be
\frac {n_X}{n_\gamma} \approx \frac{1}{m_{Pl} m_X \sigma_{ann} v}
\label{n-X-to-n-gamma} ,
\ee

where $n_X$and $n_\gamma$ are the contemporary number densities of X-particles and CMB photons respectively, 
$m_{Pl} = 1,2 \cdot 10^{19} $ GeV is the Planck mass, and $\sigma v$ is the product of the annihilation 
cross-section of $X\bar X$ by their center of mass velocity. This result is valid for a simple order of magnitude 
estimates with some numerical and logarithmic factor of order10 neglected.

For S-wave annihilation
\be 
\sigma_{ann} v  = \alpha^2 / m_X^2
\label{sigma-v}
\ee
The estimate (\ref{rho-LSP}) is obtained with $\alpha \approx 0.01$, which is typical for SUSY. If the coupling
is different and/or the annihilation is enhanced or suppressed, the result would be  evidently changed. Anyhow the
presented expressions are the conventional ones for the estimates of usual WIMPs number and energy densities.

Though there exist other mechanisms of LSP production/annihilation, which may be realized in cosmology,
nevertheless a study of alternative  cosmological models for LSPs as viable dark matter candidate can be of
interest.

Equation~(\ref{rho-LSP}) and our results obtained in Ref.~\cite{Arbuzova:2018apk} 
as well as in the present paper do not demand full supersymmetry and   are valid for any massive stable particle with the coupling strength typical to that in supersymmetry. So in what follows we will not use the abbreviation LSP for these particles but instead call them $X$-particles.
 
 
In our recent work~\cite{Arbuzova:2018apk}, we have shown that in modified $R+R^2$ cosmology the 
relative density of
LSP can be considerably smaller than that predicted in the standard scenario. This opens the window for the lightest
supersymmetric particle with the mass about 1000 TeV to be a viable dark matter candidate. The frozen number
density of massive relics is calculated 
in terms of the present day density of photons of the cosmic microwave 
background radiation, see e.g. Ref.~\cite{AD-YaZ-rev}. The relative decrease of the LSP density in $R^2$-cosmology
is related to an efficient particle production by the oscillating curvature scalar after freezing of the LSP production
and annihilation, as it is shown in our paper~\cite{Arbuzova:2018ydn}. Consequently, the number density of CMB 
photons rises and the ratio of the frozen number density of  LSP  to the number density of photons drops down.

Production of dark matter particles by the scalaron decay in a different aspect was considered also
in Ref.~\cite{Gorbunov:2010bn}.

According to our work \cite{Arbuzova:2018ydn},  the cosmological evolution in $R+R^2$ theory is 
considerably different from the standard one, based on the classical General Relativity (GR). In modified
gravity the cosmological evolution can be categorized into the following
four epochs. At first, there  was the exponential expansion (Starobinsky inflation~\cite{star-infl}),  when
the curvature scalar $R(t)$ (called scalaron) was very large and was slowly decreasing down to zero.
The next epoch began when $R(t)$ dropped 
down to zero and started to oscillate, periodically changing sign. 
The oscillations of $R$ led to particle production and this epoch
can be called Big Bang. Next, there was the transition period from the scalaron domination to the relativistic 
matter domination. Finally, after scalaron had decayed completely, we arrived to the standard cosmology 
which is governed by General Relativity. 

The frozen density of massive species strongly depends upon the probability of particle production by 
$R(t)$. In our previous papers~\cite{Arbuzova:2018apk,Arbuzova:2018ydn}, we considered the decays into
minimally coupled massless scalar particles and into massive fermions or conformally coupled 
scalars. However, as it is argued in Ref.~\cite{Gorbunov:2012ns}, the production of massless gauge 
bosons due to conformal anomaly may be significant. 
We avoided this problem assuming a version of supersymmetric model, where conformal anomaly is absent.
Here we clear out this restriction and consider freezing of massive species in the theory where the particle
production by oscillating curvature predominantly proceeds through anomalous coupling to gauge bosons. 

The paper is organized as follows. In Sec. \ref{s-cosm-evol} we summarize our results on cosmological evolution in $R^2$-gravity 
and present the known theoretical estimates of the variation of the coupling constants with changing momentum transfer.
In Sec.~\ref{s-decay-into-X} an estimate of the cosmological number density of $X$-particles created by direct decay of the
scalaron is presented. It is shown that  $X$-particles energy density would have the proper for DM value if their mass is rather small, $M_X \sim 10^7$ GeV. 
However, as shown in Sec.~\ref{X-in-plasma}, in this case the $X$-particle production by thermal processes in plasma, in turn,
would be unacceptably strong.  To avoid this crunch we assume that $X$-particles are Majorana fermions because in this
case their direct production by the scalaron is forbidden. According to the calculations in Sec.~\ref{X-in-plasma} the cosmological
density of  $X$-particles would be equal to the observed density of  DM if $M_X \sim 10^{10} $ GeV. In Sec.~\ref{s-obs} possible manifestations of
$X$-particles in cosmic rays are considered. 
 In Conclusion, the results are 
discussed and compared to the other cases studied earlier.

\section{Cosmological evolution in $R^2$ gravity  \label{s-cosm-evol}}

This section contains a condensed summary of the main results of our 
works~\cite{Arbuzova:2018apk,Arbuzova:2018ydn}. The action of the theory has the form:
\be
S_{tot} = -\frac{m_{Pl}^2}{16\pi} \int d^4 x \sqrt{-g} \left(R-\frac{R^2}{6M^2_{R}}\right)+S_m\,,
\label{S-R2-tot-1}
\ee
where $m_{Pl} = 1.2 \cdot 10^{19} $ GeV is the Planck mass, $S_m$ is a matter action. 
 Here $g$ is the determinant  of the metric tensor $g_{\mu\nu}$ taken with the signature convention 
$(+,-,-,-)$. The Riemann tensor describing the curvature of space-time is determined according to 
$R^\alpha_{\,\,\mu\beta\nu}=\partial_\beta\Gamma^\alpha_{\mu\nu}+\cdots$, 
$R_{\mu\nu} = R^\alpha_{\,\,\mu\alpha\nu}$, and
$R=g^{\mu\nu}R_{\mu\nu}$. We use here the natural system of units $\hbar=c=k_B=1$.
As we see in what follows, $M_R$ is the mass of the scalaron field. The spectrum of the temperature 
fluctuations of the cosmic microwave background radiation (CMB) demands~\cite{tegmark,Gorbunov:2010bn}:
\be
M_R = 3 \cdot 10^{13}\,\,{\rm GeV}.
\label{M-R}
\ee

We consider homogeneous and isotropic matter distribution 
with the linear equation of state: 
\be
P = w \rho,
\label{eq-state}
\ee
where $w$ is usually a constant parameter. For non-relativistic matter 
$w=0$, for relativistic matter $w=1/3$, and for the vacuum-like state $w=-1$. 

Equation of motion for the curvature  which follows from action (\ref{S-R2-tot-1}) has the form:
\begin{equation}
\ddot R - \frac{\Delta R}{a^2} +  (3H +\Gamma/2) \dot R +  M_R^2 R = - \frac{8\pi M_{R}^2}{m_{Pl}^2}(1 - 3w)\rho ,
\label{ddot-R}
\end{equation}
where $H =\dot a /a $ is the Hubble parameter and
$\Gamma$ is the total scalaron decay rate, which is determined by the dominant decay channel.
See discussion and the list of references in our works~\cite{Arbuzova:2018apk,Arbuzova:2018ydn}.
Note that our definition of $\Gamma$ in the present paper differs by factor 2 from that used in our earlier works.

The  appearance of the damping term, $\Gamma \dot R $, in this equation is a result of the back-reaction of particle production by
oscillating curvature on the curvature field. This equation has been derived in one-loop approximation in several 
papers~\cite{AD-SH,AD-KF,EA-AD-LR}. The resulting impact of particle production on 
the evolution of $R$ is described by non-local in time equation, 
which for harmonic oscillations of the source is reduced to simple liquid friction term, as given above in Eq.~(\ref{ddot-R}).
It is noteworthy that the quantum average of the energy-momentum tensor over vacuum or in external (gravitational, as in our case) 
field does not have the same value of $w$ as  for real fields, for example, vacuum expectation value of massless fields has $w=-1$ instead of $w= 1/3$.

We assume that the scalaron field is homogeneous $R=R(t)$, neglecting small perturbations
generated in the course of inflation.
After inflation is over, the scalaron field starts to oscillate as 
\be
R(t) = \frac{4 M_R \cos ( M_R t + \theta)}{t},
\label{R-of-t}
\ee
where $\theta$ is a constant phase determined by initial conditions. This and the subsequent equations are
valid in the limit $M_R t \gg 1$, but $\Gamma t \lesssim 1$.

The Hubble parameter is similar to that at the matter dominated stage (MD), but with fast oscillations around 
the MD value:
\be 
H(t) =\frac{2}{3t} \left[ 1 + \sin (M_R t + \theta) \right]
\label{H-of-t}
\ee

The cosmological energy density of matter at this period depends upon the decay width of the scalaron,
which in turn depends upon the dominant decay channel.

If there exists scalar particle minimally coupled to gravity, the decay width of scalaron into massless scalars would be:
\be
\Gamma_S = \frac{M_R^3}{24 m_{Pl}^2}.
\label{Gamma-S}
\ee
 In this case, the energy density of predominantly relativistic matter is  equal to:
\be 
\rho_{S} (t) = \frac{M_R^3}{120\pi t}  \approx 2.7\cdot 10^{-3}\,\frac{M_R^3}{ t}. 
\label{rhos-of-t}
\ee
If there are several species of massless scalars, the expressions (\ref{Gamma-S}) and (\ref{rhos-of-t})
should be multiplied by $g_S$, where $g_S$ denotes the number of species.
For massive scalar with the mass $m_s$ the width of two-body decay would be somewhat suppressed due to the phase space factor
proportional to $ \sqrt{1 - 4 m^2_s /M_R^2}$.

If scalaron predominantly decays into fermions or conformally coupled scalars the decay width vanishes
in the limit of massless final state particles and is equal to~\cite{Gorbunov:2010bn}:
\be
\Gamma_f = \frac{ M_R m_f^2 }{24 m_{Pl}^2 },
\label{Gamma-f}
\ee
where $m_f$ is the mass of fermion or conformally coupled scalar. 
The width is dominated by the heaviest final particle.
The corresponding matter density is:
\be 
\rho_{f} (t) = \frac{ M_R m_f^2}{120 \pi t}.
\label{rhof-of-t}
\ee

\begin{figure}[!htbp]
  \centering
  \begin{minipage}[b]{0.45\textwidth}
    \includegraphics[width=\textwidth]{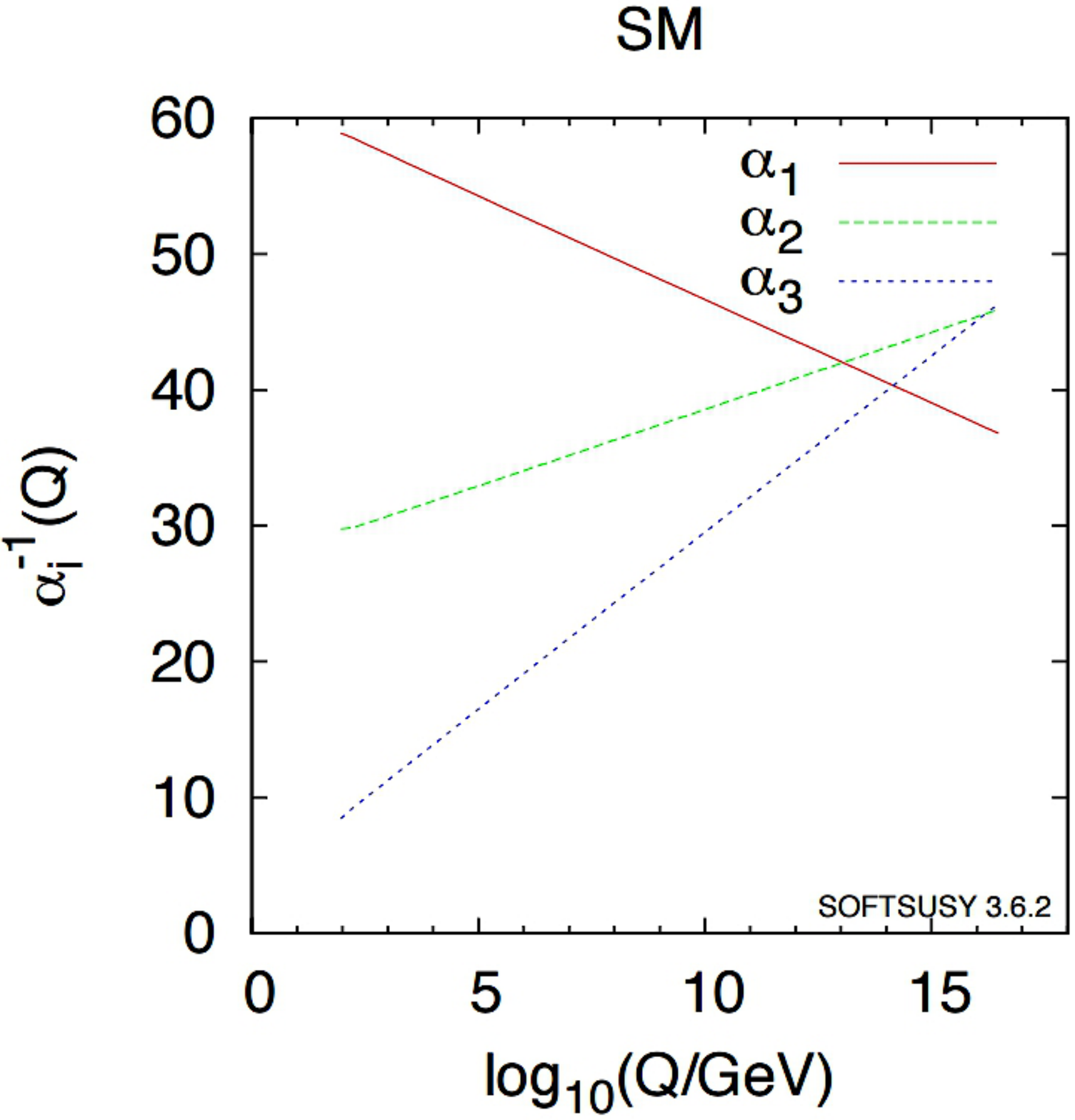}    
  \end{minipage}
  \hspace*{.1cm}
  \begin{minipage}[b]{0.45\textwidth}
    \includegraphics[width=\textwidth]{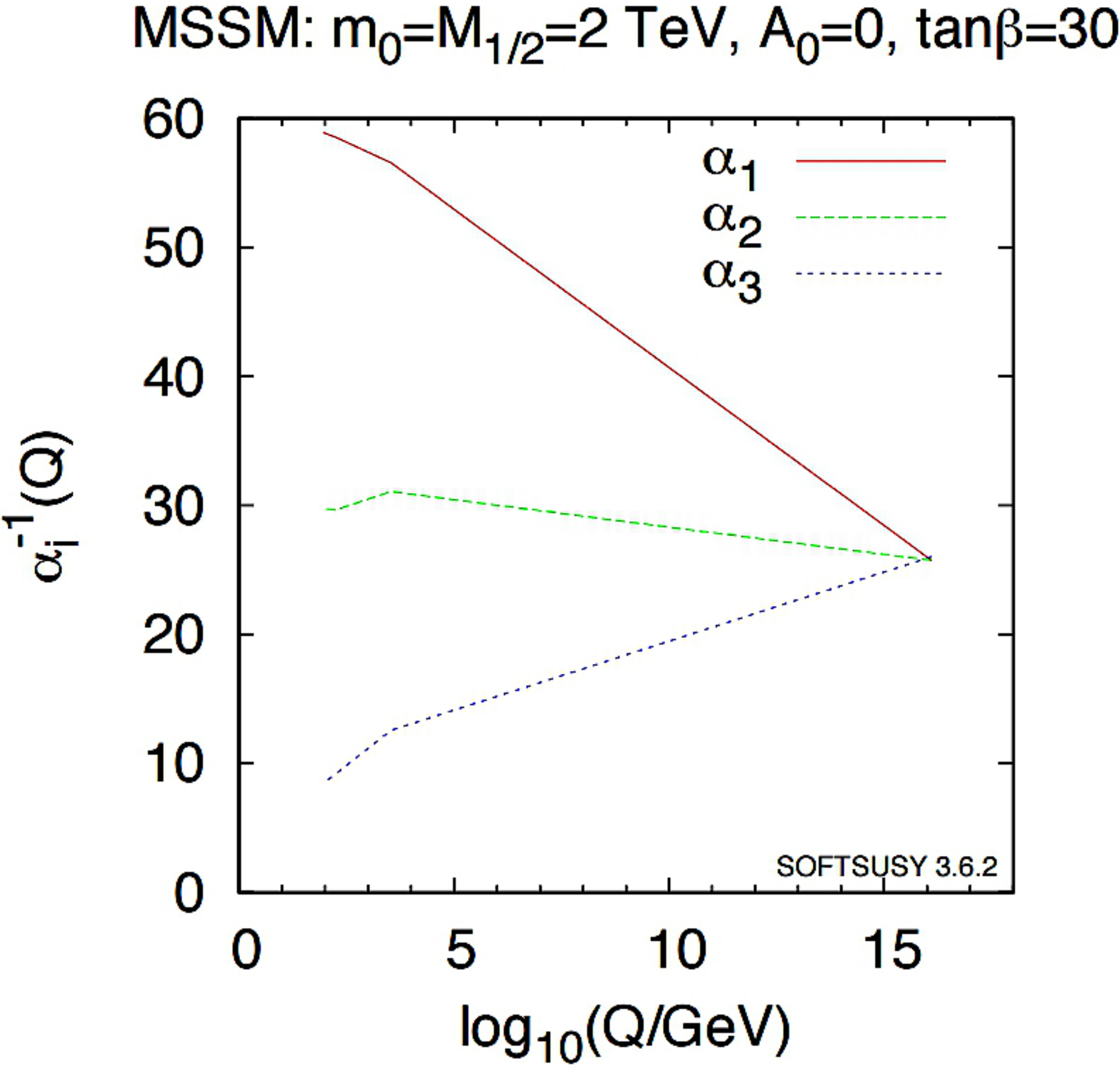}
      \end{minipage}
  \caption{Evolution of the coupling constants of $U(1)$, $SU(2)$ and  and $SU(3)$ (color) groups as function of the momentum
  transfer~\cite{PDG}. }
  \label{alpha-of-Q}
 \end{figure}

Now let us turn to the scalaron decay induced by the conformal anomaly. Production of massless gauge 
bosons by conformally flat gravitational field was first studied in Refs.~\cite{dolgov-an1,dolgov-an2} 
and applied to the problem of heating in $R^2$-inflation in Ref.~\cite{Gorbunov:2012ns}.
The scalaron decay width for this channel is equal to:
\be
\Gamma_{an} = \frac{\beta_1^2 \alpha^2 N}{96\pi^2}\,\frac{M_R^3}{m_{Pl}^2} ,
\label{Gamma-an}
\ee
where $\beta_1$ is the first coefficient of the beta-function, $N$ is the rank of the gauge group, and
$ \alpha$ is the gauge coupling constant. We take $\beta^2_1 = 49$, $N=8$. The coupling constant $\alpha$
at very high energies depends upon the theory and is, strictly speaking, 
unknown. The evolution of $\alpha$ in the minimal 
standard  model (MSM) is presented in Fig.~\ref{alpha-of-Q}, left panel, and the same in the minimal standard 
supersymmetric model (MSSM) with supersymmetry at TeV scale is presented in the right panel. We can 
conclude that at the scalaron mass scale, $Q=3\cdot 10^{13}$ GeV, $\alpha_3  \approx 0.025$ in MSM, while
in MSSM it is  $\alpha_3 \approx 0.04$.  
At $Q=10^{10}$ GeV they are $\alpha_3 \approx  0.033$ for MSM and
$\alpha_3 \approx  0.05$ for MSSM. 

The values of the running coupling constants are known to depend upon the particle spectrum. In the case 
of MSM we assumed that there exists only already known set of particles, while in MSSM there is some freedom
depending on the explicit form of the SUSY model. However the variation of the couplings related to this uncertainty
does lead to strong variation of our order of magnitude estimate of the allowed value of the mass of dark matter
particles.

Since, according to our results presented  below, supersymmetry may possibly
be realized at energies about $10^{12} -10^{13}$ GeV, the running of couplings according to MSM without
inclusion of SUSY particles is probably correct below the SUSY scale.  Recall that
for particles produced at the scalaron decay $Q=3\cdot10^{13} $ GeV, while at the universe heating temperature
after the complete decay of the scalaron it is near $10^{10}$ GeV.


So numerically the decay width is:
\be
\Gamma_{an} = 2.6\cdot 10^{-4}  \left(\frac{\alpha}{0.025}\right)^2 \, \frac{M_R^3}{m_{Pl}^2}.
\label{Gamma-an-num}
\ee 
Correspondingly the energy density of matter created by the decay into this channel would be:
\begin{equation}
\rho_{an} = \frac{\beta^2_1 \alpha^2 N}{4 \pi^2} \,\frac{M_R^3}{120 \pi t} \approx 1.65\cdot 10^{-5} \left(\frac{\alpha}{0.025}\right)^2\,\frac{M_R^3}{t} .
\label{rho-an-of-t}
\end{equation}

 It is instructive to compare the rate of the energy transferred to matter produced in 
three different cases  of the scalaron decay into minimally coupled scalars, fermions, and gauge bosons due to conformal 
anomaly with the energy density of the scalaron. To this end we need to define the energy density of the oscillation scalar  
curvature. The first term of action (\ref{S-R2-tot-1}) in the Jordan frame in the high frequency limit
can be rewritten in terms of the cosmological scale factor $a(t)$ in the way analogous to the derivation of the 
Friedmann equations performed in Ref.~\cite{SB-AD-UFN}.
For high frequency oscillations
and and large value of $M_R t$ we have found the solutions~\cite{Arbuzova:2018ydn}
\begin{equation}
H = \frac{2}{3t}\,\left[ 1 + \sin( M_R t+\theta) \right], \, \,\,R =-\frac{4M_R \cos (M_R t+\theta)}{t}.
\label{H-R}
\end{equation}
Curvature scalar is related to the Hubble parameter according to: 
\be
R = -6 \dot H - 12 H^2 \rightarrow - 6 \dot H .
\label{R-thru-H}
\ee
The last relation is valid in high frequency limit and for the oscillating parts of $H$ and $R$ which
presumably give dominant contribution to the energy density.

Keeping this in mind we can rewrite action (\ref{S-R2-tot-1}) as:
\begin{multline}
S(a) =  \frac{ 6 m_{Pl}^2}{16\pi M_R^2} \int d^4 x  a^3 \left[ M_R^2 ( \dot H + 2H^2)  + \dot H^2\right] =
\frac{3 m_{Pl}^2}{4\pi M_R^2} \int d^4 x\, a^3\, \left[ -\frac{M_R^2 H^2}{2}  + \frac{\dot H^2}{2} 
\right].
\label{S-R2-tot}
\end{multline}
The last equality is obtained through integration by parts.

Varying over the scalar field $H$ we obtain the equation of motion with the left hand side:
\be 
\ddot H + M_R^2 H = r.h.s.,
\label{ddot-H}
\ee
which is exactly the same as the r.h.s. of Eq.~(\ref{ddot-R}). This equation has the oscillating solution
multiplied by a slow function of time, such as  the presented above solution $H \sim \sin (M_R t+\theta)/t $.

Now we need to introduce canonically normalized scalar field $\Phi$ linearly connected 
with $H$ for which the kinetic term in the Lagrangian is equal to $(\partial \Phi)^2/2$:
\be
\Phi = \sqrt{\frac{3}{4 \pi}}\, \frac{m_{Pl}}{M_R}\,H.
\label{Phi}
\ee
According to the standard prescription the energy density of the scalar field $\Phi$ is 
\be
\rho_{\Phi} = \frac{\dot\Phi^2 +M_R^2 \Phi^2}{2}. 
\label{rho-Phi}
\ee
 Since, according to Eq.~(\ref{R-thru-H}), in high frequency limit 
$ R =  -6 \dot H \sim  M_R \cos (M_R t) $  Eq.~(\ref{rho-Phi}) can be identically rewritten in terms of R as 
\be
\rho_R  
= \frac{m_{Pl}^2 (\dot R^2 + M_R^2 R^2)}{96 \pi M_R^4} =
\frac{m_{Pl}^2}{6\pi t^2} ,
\label{rho-R}
\ee
where expression (\ref{R-of-t}) has been used. This result coincides with the expression for the total cosmological
energy density in spatially flat matter dominated universe. 
This agreement confirms the validity of our approach.

The presented equations are valid if the energy density of matter remains smaller than the energy density 
of the scalaron until it decays.
Comparing Eqs.~(\ref{rhos-of-t}), (\ref{rhof-of-t}), and (\ref{rho-an-of-t}) with (\ref{rho-R}) we find that in all the cases
$t_{cr} \Gamma = 5/6$, where $t_{cr}$ is the time when the matter energy density, formally taken, is equal to the
scalaron energy density. So the used above equations are not unreasonable. 
The scalaron completely decays at $t = 1/\Gamma$ (up to log-correction) and the cosmology turns into the usual Friedmann one 
governed by the equations of General Relativity (GR). Before that moment the universe expansion was 
dominated by the scalaron.

If the primeval plasma is thermalized, the following  relation between the cosmological time and the
temperature is valid:
\be
\rho_{an} = 2.6\cdot 10^{-2} \alpha_R ^2\, \frac{M_R^3}{t} = \frac{\pi^2 g_*}{30}\, T^4 ,
\label{time-temp-1}
\ee
where subindex $R$ at $\alpha_R$  means that the coupling is taken at the energies equal to the scalaron mass, since the
energy influx to the plasma is supplied by the scalaron decay,
and $g_* \approx 100$ is the number of relativistic species.
Consequently,
\be
t T^4  = \frac{0.78 }{\pi^2  g_*}\,  \alpha_R^2 M_R^3 \equiv C \equiv C_0 M_R^3 
\label{t-T4}
\ee
with $C_0 = 5\cdot 10^{-7} (\alpha_R/0.025)^2 $.

Thermal equilibrium is established if the reaction rate 
is  larger than the Hubble expansion rate $H = 2/(3t)$.
The reaction rate is determined by the cross-section of two-body reactions between
relativistic particles. The typical value of this cross-section  at high energies, $E\gg m$, is~\cite{AB}:
\be
\sigma_{rel} = \frac{4\pi  \beta \alpha^2}{s} \left( \ln  \frac{s}{4 m^2} +1 \right)  ,
\label{sigma-rel}
\ee
where $\beta \sim10$ is the number of the open reaction channels and
$s = (p_1 +p_2)^2 = 4E^2$ is the total energy of the scattering particles in their center-of-mass frame, where $E$ is the
energy of an individual particle.

Hence the reaction rate is 
\be
\Gamma_{rel} \equiv \frac{\dot n}{n} = \langle \sigma_{rel} v n _{rel}\rangle,
\label{Gamma-rel}
\ee 
where angular brackets mean averaging over thermal bath with temperature $T$, $n_{rel} \approx 0.1 g_* T^3 $ (we do not distinguish between
bosons and fermions in the expression), $v=1$ is the particle velocity in the center-of-mass system.
We perform thermal averaging naively taking $E = T$ in all expressions so $s \rightarrow 4T^2$, instead of  $m^2 $ we substitute the particle
thermal mass in plasma, i.e. $m^2 \rightarrow 4 \pi \alpha T^2 /3 $~\cite{m-of-T1,m-of-T2,m-of-T3}.
Correspondingly we arrive to the following thermal equilibrium condition:
 \be
 {\frac{3}{2} \, t \Gamma_{rel} = 
0.15 \pi  \alpha^2 \beta g_*}\,\left(\ln \frac{3}{4 \pi \alpha} +1 \right) Tt  > 1.
\label{equil-1}
 \ee

Using Eq.~(\ref{t-T4}),
we find that equilibrium is established at the temperatures below 
\begin{equation}
T_{eq}  = \left( 
0.15 \pi \alpha^2 \beta g_* C_0 \right)^{1/3} M_R  = 9.2 \cdot 10^{-3} M_R.
\label{T-eq}
\end{equation}
Here we took $\alpha_R  = 0.025$ and $\alpha = 0.033$. 

The time corresponding to this temperature  is equal to
\be
t_{eq} = C/T^4_{eq} \approx 70 M_R^{-1},
\label{t-eq}
\ee
where $C$ is defined in Eq.~(\ref{t-T4}).
Hence $M_R t_{eq} \gg 1$, which is sufficiently long time for efficient particle production.

Another essential temperature for our consideration, is the temperature of the universe heating, when
scalaron essentially  decayed and the expansion regime turned to the conventional GR one. This temperature 
is determined by the scalaron energy density at the moment $t = 1/\Gamma_{an}$:
\be 
\rho_R = \frac{m^2_{Pl} \Gamma^2_{an}}{6\pi} = \frac{\pi^2 g_*}{30} \,T_h^4,
\label{T-h1}
\ee
so 
\begin{equation}
T_h =   3.2\cdot 10^{-3} \sqrt{M_R/m_{Pl}}\, M_R = 5.1 \cdot 10^{-6} M_R .
\label{T-h}
\end{equation}


\section{X-particle production through the scalaron decay \label{s-decay-into-X}}

There are two possible channels to produce massive stable X-particles: first, directly through the scalaron decay
into a pair $X \bar X$ and another by inverse annihilation of relativistic particles in plasma. 

Firstly, let us consider the scalaron decay. The probability of the scalaron decay into a pair
of fermions  is determined by decay width~(\ref{Gamma-f}) with  the substitution $M_X$ instead of $m_f$:
\be
\Gamma_X = \frac{ M_R M_X^2 }{24 m_{Pl}^2 }.
\label{Gamma-X}
\ee
The branching ratio of this decay is equal to:
\be
BR (R \rightarrow X \bar X) = \frac{ \Gamma_X }{ \Gamma_{an}} \approx 1.6\cdot 10^2  \left(\frac{M_X}{M_R}\right)^2 .  
\label{BR}
\ee
The number density of X-particles created by the scalaron decay only, 
but not by inverse annihilation of
relativistic particles in plasma, 
is governed by the equation:
\be
\dot n_X + 3H n_X = \Gamma_X n_R, 
\label{dot-nX-1}
\ee
where $\Gamma_X$ is given by Eq.~(\ref{Gamma-X}), $n_R = \rho_R/ M_R $, and $\rho_R $ is defined in
Eq.~(\ref{rho-R}). So Eq.~(\ref{dot-nX-1}) turns into
\be
\dot n_X + 3H n_X =\frac{1}{24} \frac{ M_X^2}{6\pi t^2}\, .
\label{dot-nX-2}
\ee
It is solved as
\be
n_X = \frac{ 1}{144\pi }\, \frac{M_X^2}{t}.
\label{nX-1}
\ee
The equations presented above are valid if the inverse decay of the scalaron can be neglected. This approximation is true if the
produced particles are quickly thermalized down to the temperatures much smaller than the scalaron mass.

We are interested in the ratio of $n_X$ to the number density of relativistic species at the moment of the complete
scalaron decay when the temperature dropped down to $T_h$ (\ref{T-h}) and after which the universe came to
the conventional Friedmann cosmology and the ratio $n_X/n_{rel}$ remained constant to the present time. This ratio
is equal to:
\begin{multline}
F \equiv  \frac{n_X}{n_{rel}} |_{T=T_h}=\left[ \frac{0.04M_X^2 }{6\pi t_h }\right] \times \left[\frac{\pi^2 g_* T_h^4}{90 T_h} \right]^{-1} \\=
2.3\cdot 10^{-3} \left(\frac{0.025}{\alpha_R}\right)^2 \left(\frac{M_X}{M_R}\right)^2.
\label{nX-to-nrel}
\end{multline}
Consequently, the energy density of $X$-particles in the present day universe would be:
\be
\rho_X^{(0)} = 412/{\rm cm^{-3}} M_X F = \rho_{DM} \approx 1 {\rm keV/cm}^3 .
\label{rhoX-0}
\ee
The last approximate equality in the r.h.s. is the condition that the energy density of X-particles is equal to
the observed energy density of dark matter. 

From this condition it follows that $M_X \approx 10^7$ GeV. For larger masses  $\rho_X^{(0)} $ would be
unacceptably  larger than $\rho_{DM}$. On the other hand, for such a small, or smaller $M_X$, the probability of X-particle
production through the inverse annihilation would be too strong and would again lead to very large
energy density of $X$-particles, see the following section.

A possible way out of this ``catch-22" is to find a mechanism to suppress the scalaron decay into a pair of
X-particles. And it does exist. If X-particles are Majorana fermions, then 
in this case particles and antiparticles are identical and so they must be in antisymmetric state. Thus the decay of 
a scalar field, scalaron, into a pair of identical fermions is forbidden, since the scalaron can produce a pair of identical particles 
in symmetric state only.

\section{Production of X-particles in thermal plasma \label{X-in-plasma}}

Here we turn to the X-production through the inverse annihilation of relativistic particles in the
thermal plasma. The number density $n_X$ is governed by the Zeldovich equation:
\be
\dot n_X + 3H n_X = \langle \sigma_{ann} v \rangle \left(n_{eq}^2 - n_X^2 \right) ,
\label{dot-nX}
\ee
where $ \langle \sigma_{ann} v \rangle$ is the thermally averaged annihilation cross-section of X-particles
and $n_{eq}$ is their equilibrium number density.

This equation was originally derived by Zeldovich in 1965~\cite{Zeldovic:1965}, and in 1977 it was applied to freezing
of  massive stable neutrinos in the papers~\cite{LW,VDZ}. After that it was unjustly named as Lee-Weinberg 
equation.

The thermally averaged annihilation cross-section of non-relativistic X-particles, 
which enters Eq.~(\ref{dot-nX}), for our case can be taken as
\be
\langle \sigma_{ann} v \rangle = \frac{\pi \alpha^2 \beta_{ann}}{M_X^2} \,\frac{T}{M_X},
\label{sigma-v}
\ee
where the last factor came from thermal averaging  
of the velocity squared of X-particles, equal to 
$\langle v^2 \rangle = T/M_X$, which appears because
the annihilation of Majorana fermions proceeds in P-wave. We take the coupling constant at  the energy scale around $M_X$ 
equal to $\alpha = 0.033$ and the number of the annihilation channels $\beta_{ann} = 10$. This expression is only an order of magnitude. The exact form depends upon particle spins, the form of the interaction, and may contain the statistical factor $1/n!$, if there
participate $n$  identical particles. In what follows we neglect these subtleties.

The equilibrium distribution of non-relativistic X-particles has the form:
\begin{multline}
n_{eq} = { g_s} \left(\frac{M_X T}{2 \pi}\right)^{3/2}\,\exp\left(-\frac{M_X}{T} \right)\\
={ g_s}\, M_X^3 \left( 2 \pi y \right)^{-3/2} \exp (-y),
\label{n-eq}
\end{multline}
where $y = M_X /T$ and  $g_s$ is the number of spin states of X-particles.
The non-relativistic approximation is justified if $M_X > T_{eq} \approx M_R/100 = 3\cdot10^{11}$GeV, see Eq.~(\ref{T-eq}).


Equation (\ref{dot-nX}) will be solved with the initial condition $n_X(t_{in}) = 0$.   This condition is essentially 
different from the solution of this equation in the canonical case, when it is assumed that 
initially $n_X = n_{eq}$ 
and in the course of the evolution $n_X$ becomes much larger than $n_{eq}$, reaching the so called frozen
density. As we see in what follows, for certain values of the parameters
the similar situation can be realized, when $n_X$ approaches the equilibrium value and freezes at much larger 
value. The other limit when $n_X$ always remains smaller than $n_{eq}$ is also possible. 

For better insight   into the problem we first make simple analytic estimates of the solution when $n_X \ll n_{eq}$ 
and after that solve exact Eq.~(\ref{dot-nX}) numerically.

In the limit $n_X \ll n_{eq}$ Eq.~(\ref{dot-nX}) is trivially integrated:

\begin{multline}
n_{X0} (y)=\frac{4\pi \alpha ^2 \beta_{ann} g_s^2}{(2\pi)^3} 
\frac{C_0 M_R^3}{y^8} \int_{y_{in}}^{y} dy_1 y_1^7 e^{-2y_1} \\
=
5\cdot 10^{-7}\,\frac{ \alpha ^2 \beta_{ann} g_s^2}{2\pi^2}\,  \left( \frac{\alpha_R}{0.025}\right)^2 M_R^3 
\int_{y_{in}/y}^{1} dz z^7 e^{-2 z y }
\label{nX-sol12}
\end{multline}
{where the subindex ``0" means that the solution is valid for $n_X \ll n_{eq}$, 
$y = M_X/T$ and we have used Eq.~ (\ref{t-T4}) and the expression for $C_0$ below this equation.

For the initial temperature we take $T_{in} = 6 \cdot 10^{-3} M_R$, 
according to Eq.~(\ref{T-eq}),
and $T_{fin} = T_h = 5.1\cdot  10^{-6} M_R$ (\ref{T-h}). Correspondingly $y_{in} = 1.7\cdot 10^2 M_X/M_R $, and
$y_{fin} = 2\cdot10^5 M_X/M_R$ and so $y_{fin}/y_{in}\approx 10^3$. 

To check validity of this solution we have to compare $n_{X0} (y)$ to $n_{eq}$ (\ref{n-eq}):
\begin{multline}
F_2 (y) \equiv \frac{n_{X0} (y)}{n_{eq}}=
5\sqrt{2}\cdot 10^{-7} \,\frac{\alpha^2 \beta_{ann} g_s}{\sqrt{\pi}} \, \\  \left(\frac{\alpha_R}{0.025} \right)^2  e^y y^{3/2} 
\int_{y_{in}/y}^1 dz z^7 e^{-2z y}  \nonumber \\
=8.7 \cdot 10^{-9} \left(\frac{M_R}{M_X}\right)^3
 e^y y^{3/2} \int_{y_{in}/y}^1 dz z^7 e^{-2z y} , 
\label{F2-of-y}
\end{multline}
where we have taken $g_s = 2$, $\beta_{ann} =10$ and lastly, according to the line below
Eq.~(\ref{T-eq}), $\alpha = 0.033$ and $\alpha_R = 0.025$.

The ratio $F_2 (y)$ is depicted in Figs.~\ref{fig-F2-01-5},  \ref{fig-F2-20-50} as function of $y$ for different 
values of $y_{in}$. The ratio remains smaller than unity for sufficiently small $y < y_{max} = 50-150$ depending upon
$y_{in}$. If $y_{fin} < y_{max}$, the assumption $n_X \ll n_{eq}$ is justified and the solution~(\ref{nX-sol12}) is a good 
approximation to the exact solution. In the opposite case, when $y_{fin} > y_{max}$, we have to solve Eq.~(\ref{dot-nX})
numerically.

\begin{figure}[!htbp]
  \centering
  \begin{minipage}[b]{0.45\textwidth}
    \includegraphics[width=\textwidth]{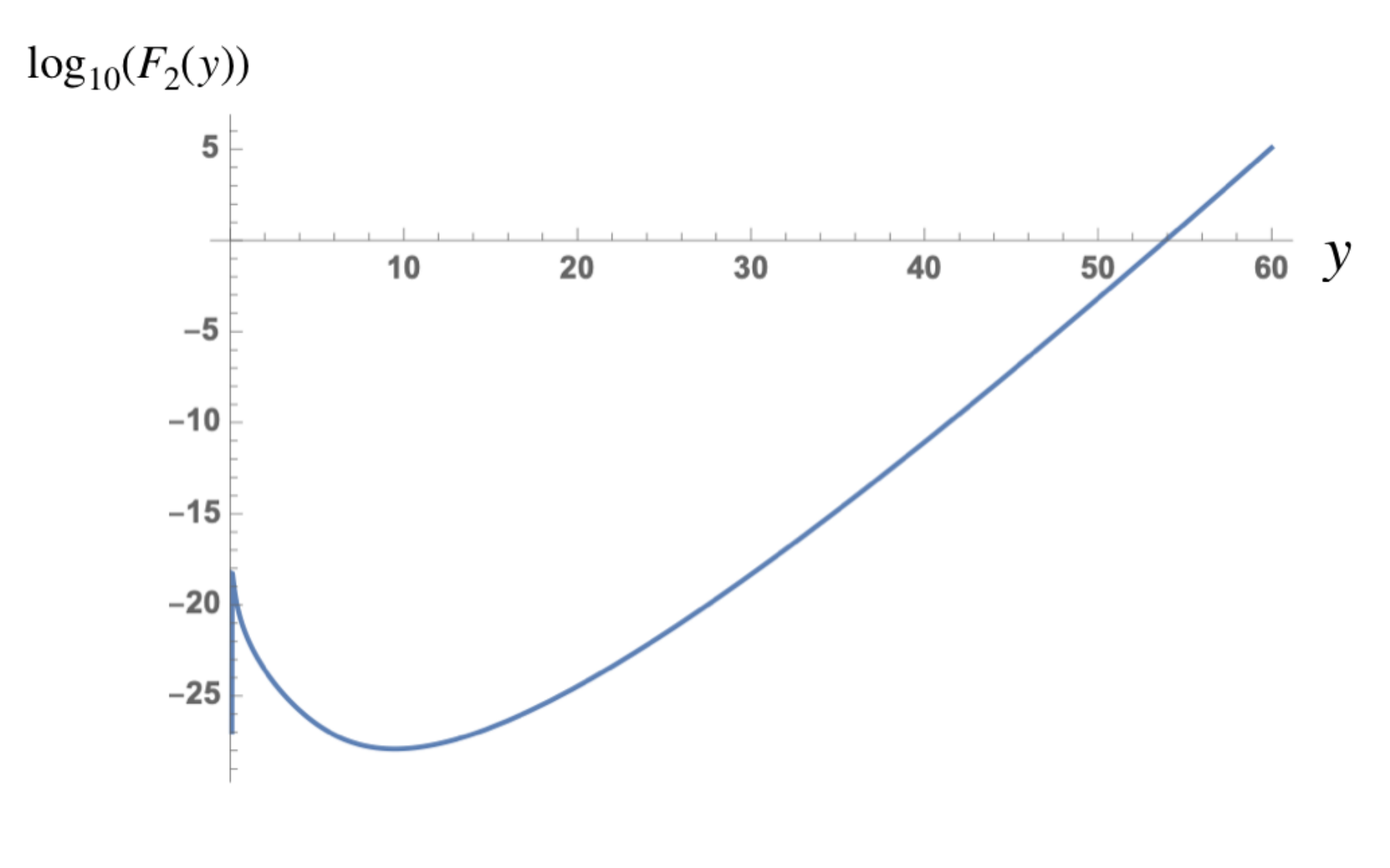}    
  \end{minipage}
  \hspace*{.1cm}
  \begin{minipage}[b]{0.45\textwidth}
    \includegraphics[width=\textwidth]{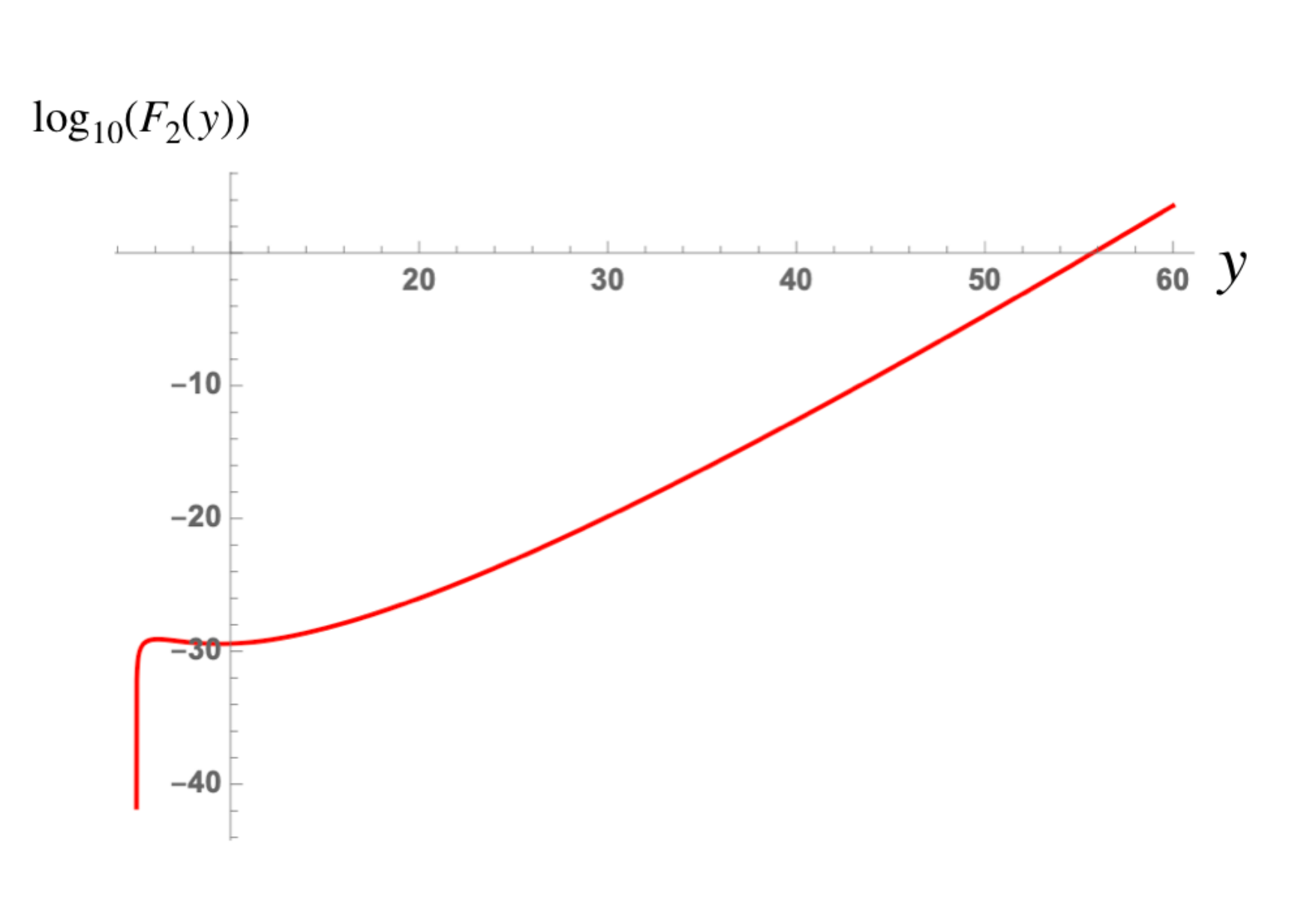}
      \end{minipage}
  \caption{ Log of ratio of the calculated number density of $X$-particles to the equilibrium number 
  density~(\ref{F2-of-y})  calculated in the limit $n_X \ll n_{eq}$; left panel: $y_{in} = 0.1$ 
  and right panel: $y_{in} = 5$.   }
    \label{fig-F2-01-5}
 \end{figure}

\begin{figure}[!htbp]
  \centering
  \begin{minipage}[b]{0.45\textwidth}
    \includegraphics[width=\textwidth]{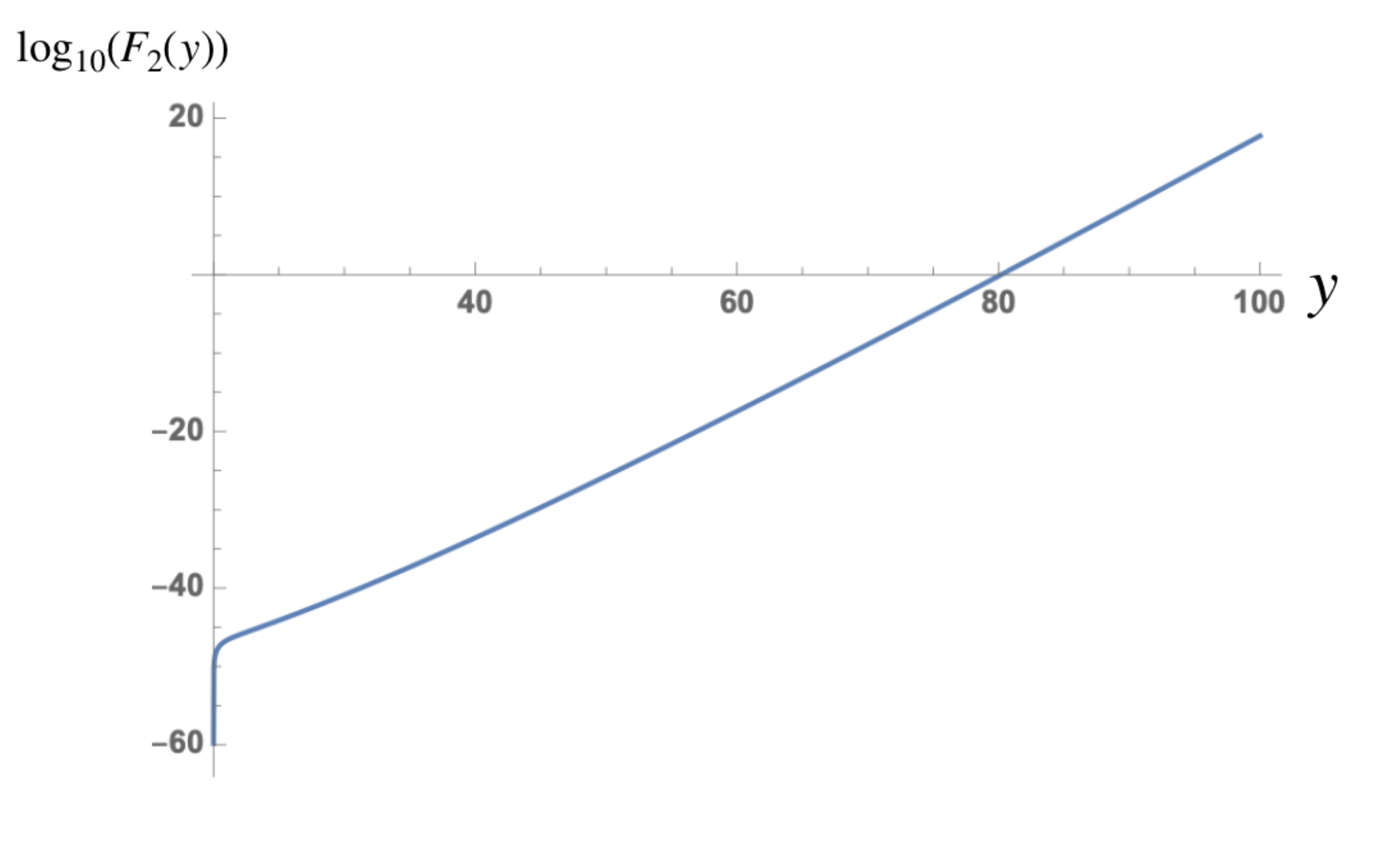}    
  \end{minipage}
  \hspace*{.1cm}
  \begin{minipage}[b]{0.45\textwidth}
    \includegraphics[width=\textwidth]{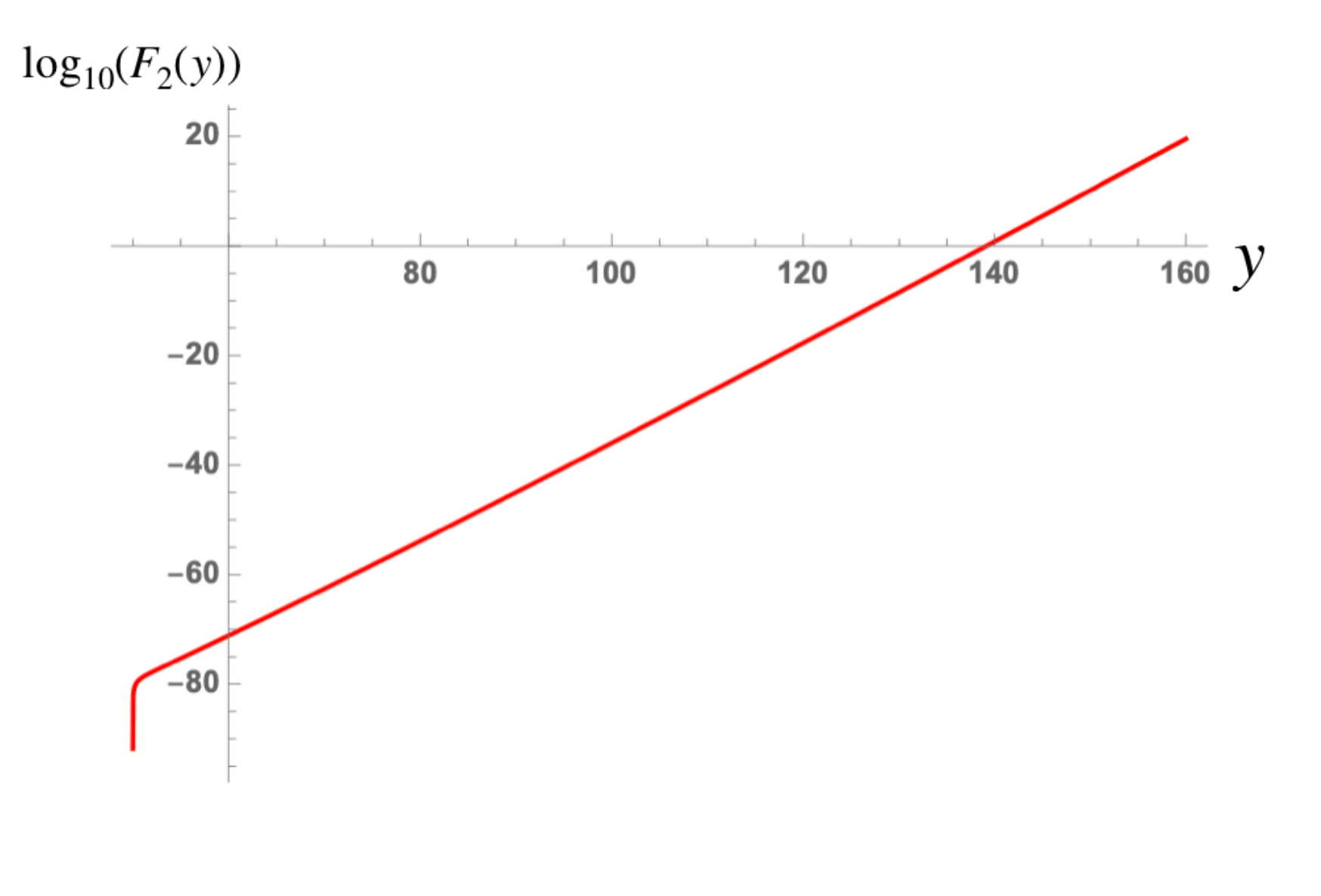}
      \end{minipage}
  \caption{ Log of the ratio of the calculated number density of $X$-particles to the equilibrium number 
  density~(\ref{F2-of-y})  calculated in the limit $n_X \ll n_{eq}$; left panel: $y_{in} = 20$ 
  and right panel: $y_{in} = 50$.   }
    \label{fig-F2-20-50}
 \end{figure}


To solve the equation~(\ref{dot-nX})  it is convenient to introduce the new function according to: 
\be
n_X = {g_s} \left( \frac{a_{in}}{a(t)} \right)^3 M_X^3 z(t) = { g_s} M_X^3 \left(\frac{T_{in}}{T}\right)^{-8} z,
\label{nX-z}
\ee
where $a(t)$ is the cosmological scale scale factor and $a_{in}$ is its initial value at some time $t=t_{in}$, when
$X$-particles became non-relativistic.
In terms of $z$, equation~(\ref{dot-nX}) is reduced to:
\be
\dot z = \langle \sigma_{ann} v \rangle M_X^3 \left( \frac{a_{in}}{a} \right)^3 \left(z_{eq}^2 - z^2 \right) ,
\label{dot-s}
\ee
Next, let us change the variables from $t$ to $y=M_X/T$. Evidently $\dot y = -y (\dot T/T)$.
Using time-temperature relation  (\ref{t-T4}), we find
\be 
\frac{d z}{d t} = \frac{M_X^4}{4  C_0 M_R^3\, y^3}\, \frac{dz}{dy}. 
\label{ds-dy-1}
\ee
Keeping in mind that
\be
\left(\frac{a_{in}}{a}\right)^3 = \left(\frac{t_{in}}{t}\right)^2 = \left(\frac{y_{in}}{y}\right)^8,
\label{ain-over-a}
\ee
we find finally:
\begin{equation}
\frac{dz}{dy} = 4\pi\, { g_s} C_0  \alpha^2 \beta_{ann}\, \mu^3
\,\frac{y_{in}^8}{y^6}\,  \left( \frac{y^{13}}{8\pi^3 y_{in}^{16} e^{2y}} - z^2 \right) 
\label{dz-dy-2} ,
\end{equation}
where $\mu = M_R / M_X$.

With the chosen above values of $\alpha_R$ and $\alpha$, see the discussion after 
Eq.~(\ref{sigma-v}), we find that the value of the coefficient in the 
r.h.s. of Eq.~(\ref{dz-dy-2}) is 
$4\pi { g_s} C_0  \alpha^2 \beta_{ann} = 1.4\cdot 10^{-7}$.

Numerical solution of this equation indicates that $z(y)$ tends asymptotically at large $y$ to a constant 
value $z_{asym}$. The energy density of X-particles is expressed through $z_{asym}$ as follows.
We assume that below $T=T_h$ the ratio of number density of X-particles to the number density of relativistic 
particles remains constant and hence is equal to the ratio $n_X/n_{CMB}$ at the preset time, 
where $n_{CMB} = 412/{\rm cm}^3$ is the contemporary number density of photons in cosmic microwave background  
radiation. The number density of $X$-particles is expressed through $z$  according to Eq.~(\ref{nX-z}). Thus the
asymptotic ratio of the number densities of X to the number density of relativistic particles is
\begin{equation}
F_{asym} =\frac{n_X (T_h)}{n_{rel} (T_h)}  = 
\left[ M_X^3 \left(y_{in}/y_h\right)^8  z_{asym}\right] \cdot
 \left[ \pi^2 g_* T_h^3/90 \right]^{-1}.
\label{nX-asym}
\end{equation}


We assume that $y_{in} \approx 10^{2}/\mu$, 
$y_{fin} = y_h= 2\cdot 10^{5}/\mu$, according to the discussion after Eq.(\ref{nX-sol12}), and so 
$y_{fin}/y_{in}\approx 2 \cdot 10^3$. Hence the energy density of X-particles today would be equal to:
\begin{equation}
\rho_X^{(0)} = (412/{\rm cm}^3) M_X  F_{asym} = 
3 \cdot 10^9 \mu^{-4}
z_{asym} \,\frac{ {\rm keV}}{{\rm cm}^3},
\label{rhoX-z}
\end{equation}
where $z_{asym}$ is the asymptotic value of $z(y) $ at large $y$ but still smaller than $y_{h}$. The value of $z_{asym}$ can be
found from the numerical solution of Eq.~(\ref{dz-dy-2}). However, the solution demonstrates surprising feature: its derivative 
changes sign at $y \lesssim 10$, when $n_X \ll n_{eq}$, as it is seen from the value of $F_2$ presented in 
Figs.~\ref{fig-F2-01-5} and \ref{fig-F2-20-50}. Probably this evidently incorrect  result for $z(y)$ originated from a very small 
coefficient in front of the brackets in Eq.~(\ref{dz-dy-2}).

The problem can be avoided if we introduce the new function $u(y)$ according to:
\be
z(y)  = \frac{u(y)}
{(2\pi)^{3/2} y_{in}^{3/2} \exp (y_{in})} .
\ee
In terms of $u(y)$ kinetic equation takes the form:
\begin{equation}
\frac{du}{dy} = \frac{1.4\cdot 10^{-7}\, \mu^3\,y_{in}^{13/2}}{(2\pi)^{3/2} y^6 \,\exp(y_{in})} \,\left[
\left(\frac{y}{y_{in}}\right)^{13}\, e^{2(y_{in} - y)} - u^2
\right] .
\label{du-dz} 
\end{equation}
The numerical solution of this equation does not show any pathological features and may be trusted, so we express the 
contemporary energy of dark matter made of stable $X$-particles through the asymptotic value of $u(y)$ as 
\be
\rho_X^{(0)} = 
\frac{ 3 \cdot 10^9 \mu^{-4}  u_{asym}}
{(2\pi)^{3/2} y_{in}^{3/2} \exp (y_{in})} 
\,\frac{ {\rm keV}}{\rm{cm}^3} .
\label{rho-X-u}
\ee
Remind that $y_{in} \approx 100/\mu$ and presumably $\mu > 1$.

The asymptotic value $u_{asym}$ is found from the numerical solution of Eq.(\ref{du-dz}) and is depicted in Figs.~\ref{u-of-y1} and \ref{u-of-y2} 
for different values of $\mu$.

\begin{figure}[!htbp]
  \centering
  \begin{minipage}[b]{0.45\textwidth}
    \includegraphics[width=\textwidth]{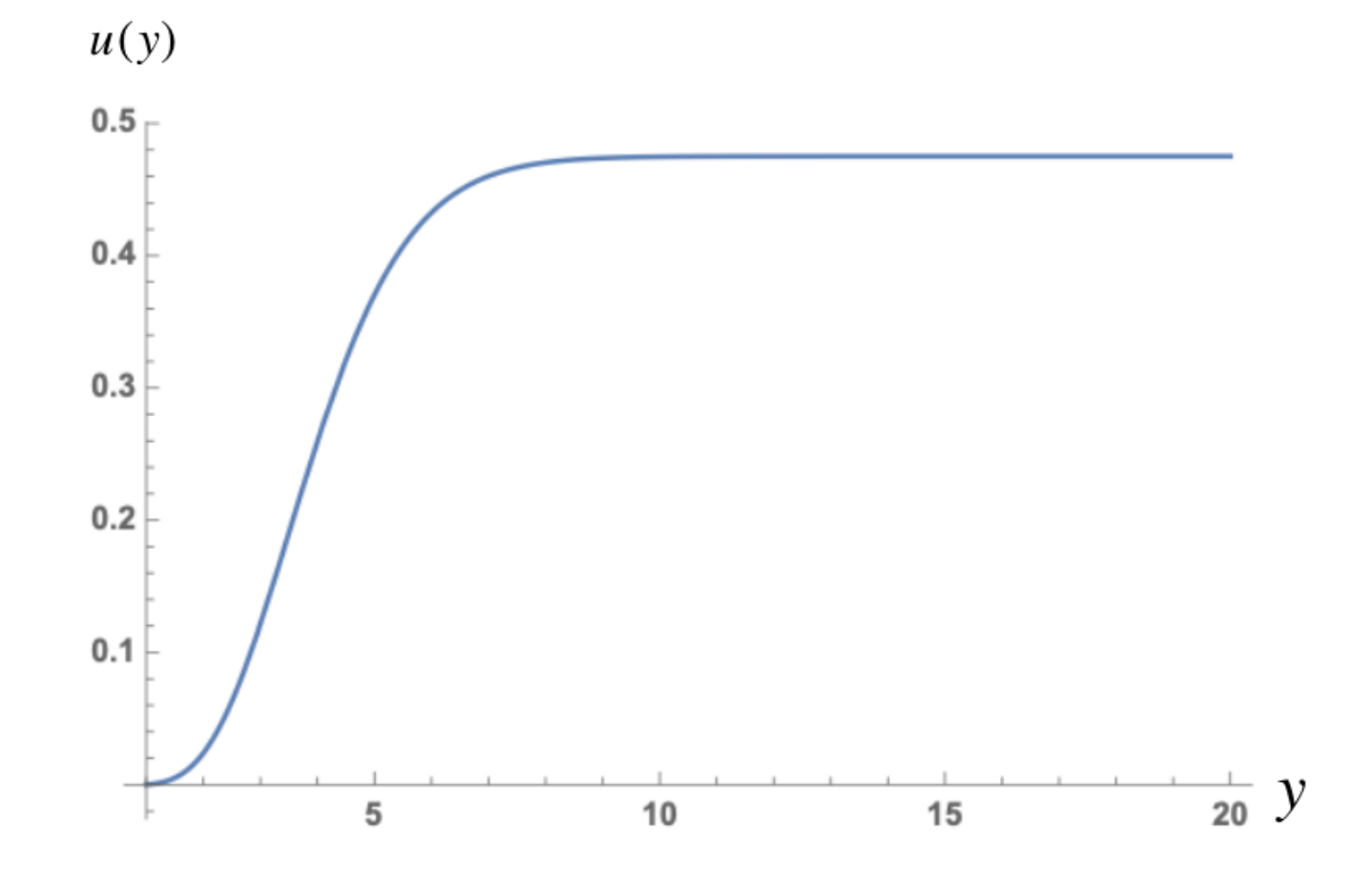}    
  \end{minipage}
  \hspace*{.1cm}
  \begin{minipage}[b]{0.45\textwidth}
    \includegraphics[width=\textwidth]{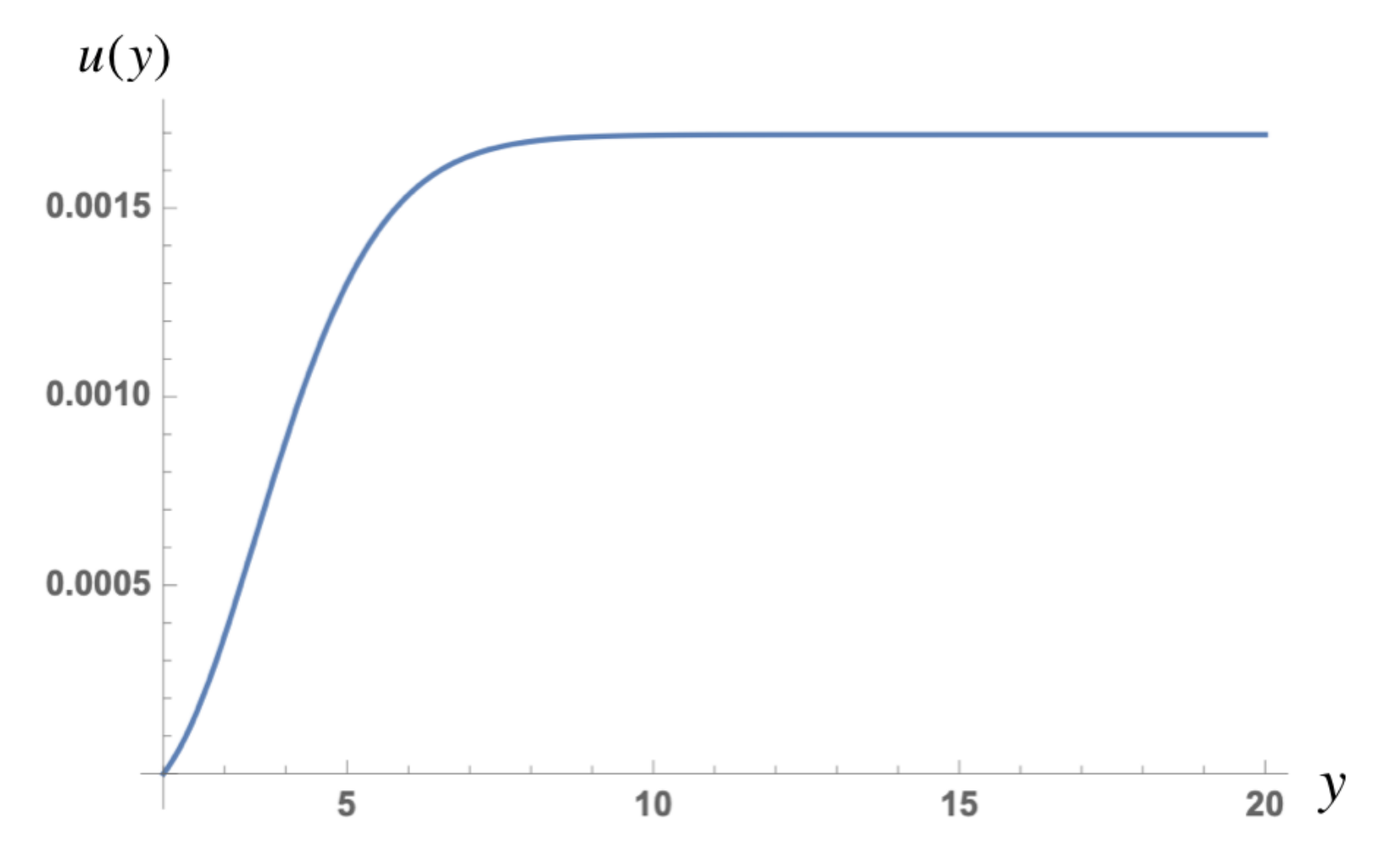}
      \end{minipage}
  \caption{Evolution of $u(y)$  for $\mu =100$ (left) and $\mu =50$ (right).}
  \label{u-of-y1}
 \end{figure}

\begin{figure}[!htbp]
  \centering
  \begin{minipage}[b]{0.45\textwidth}
    \includegraphics[width=\textwidth]{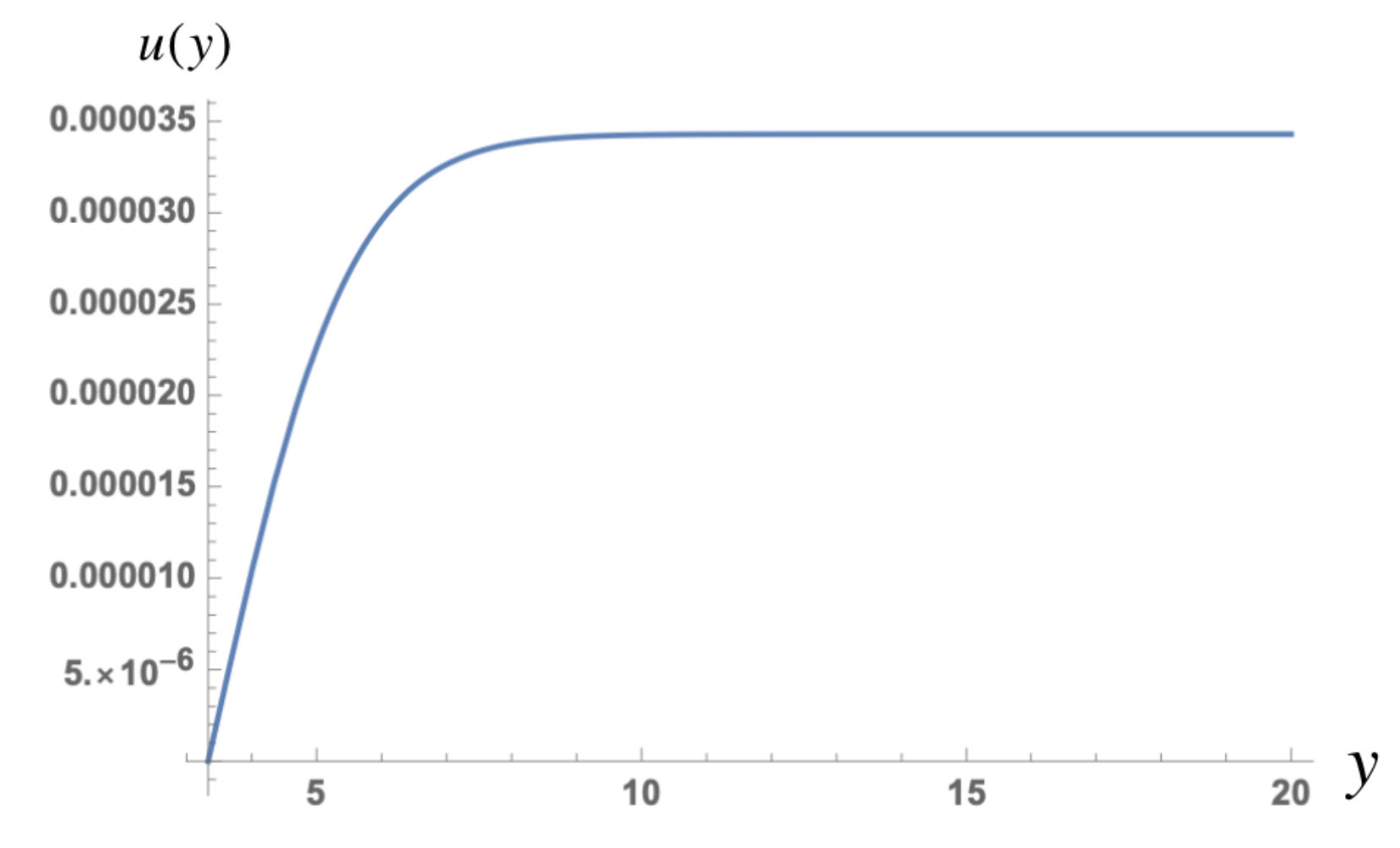}    
  \end{minipage}
  \hspace*{.1cm}
  \begin{minipage}[b]{0.45\textwidth}
    \includegraphics[width=\textwidth]{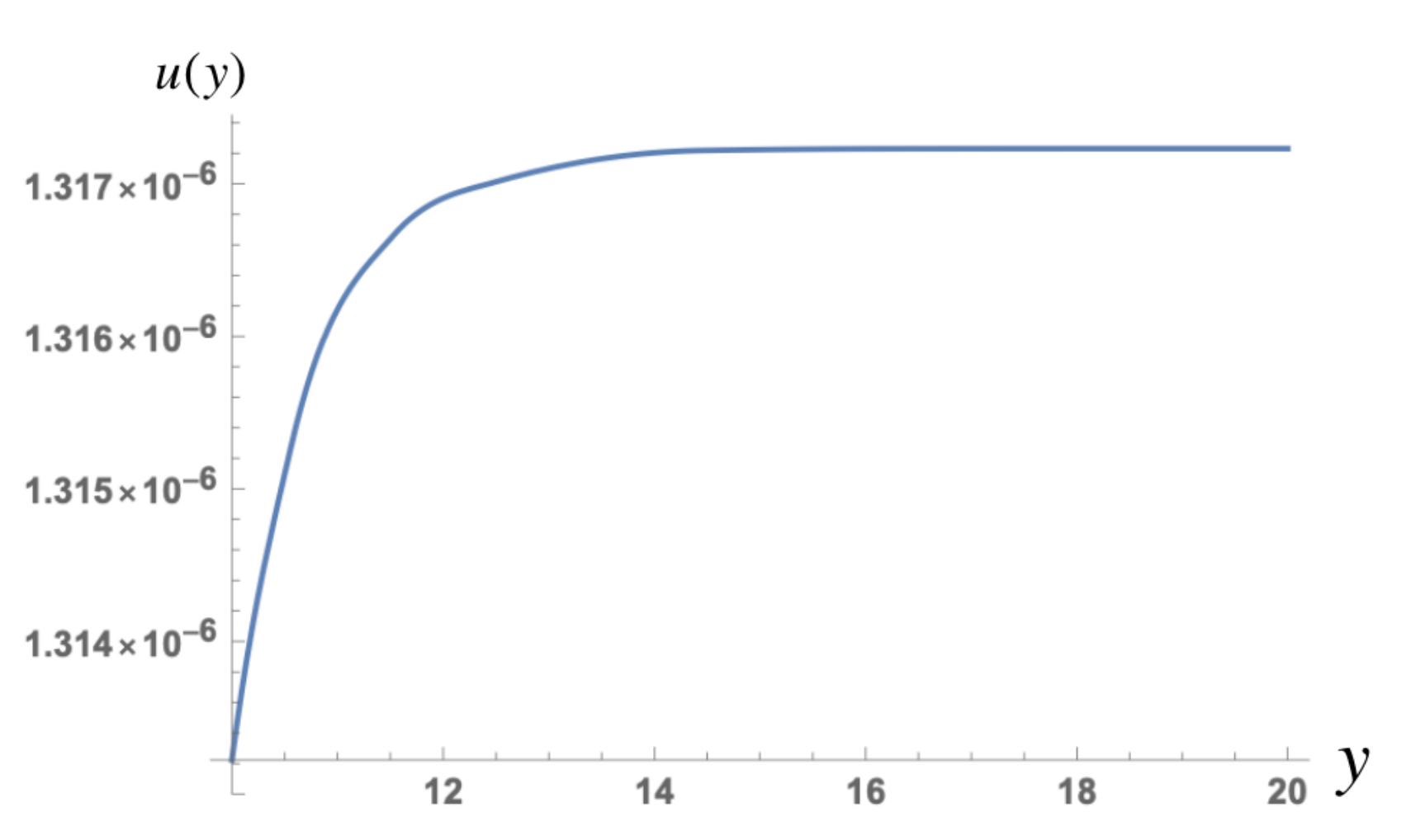}
      \end{minipage}
  \caption{Evolution of $u(y)$  for $\mu =30$ (left) and $\mu =20$ (right).
  }
    \label{u-of-y2}
 \end{figure}

\begin{figure}[!htbp]
  \centering
 \includegraphics[scale=0.5]{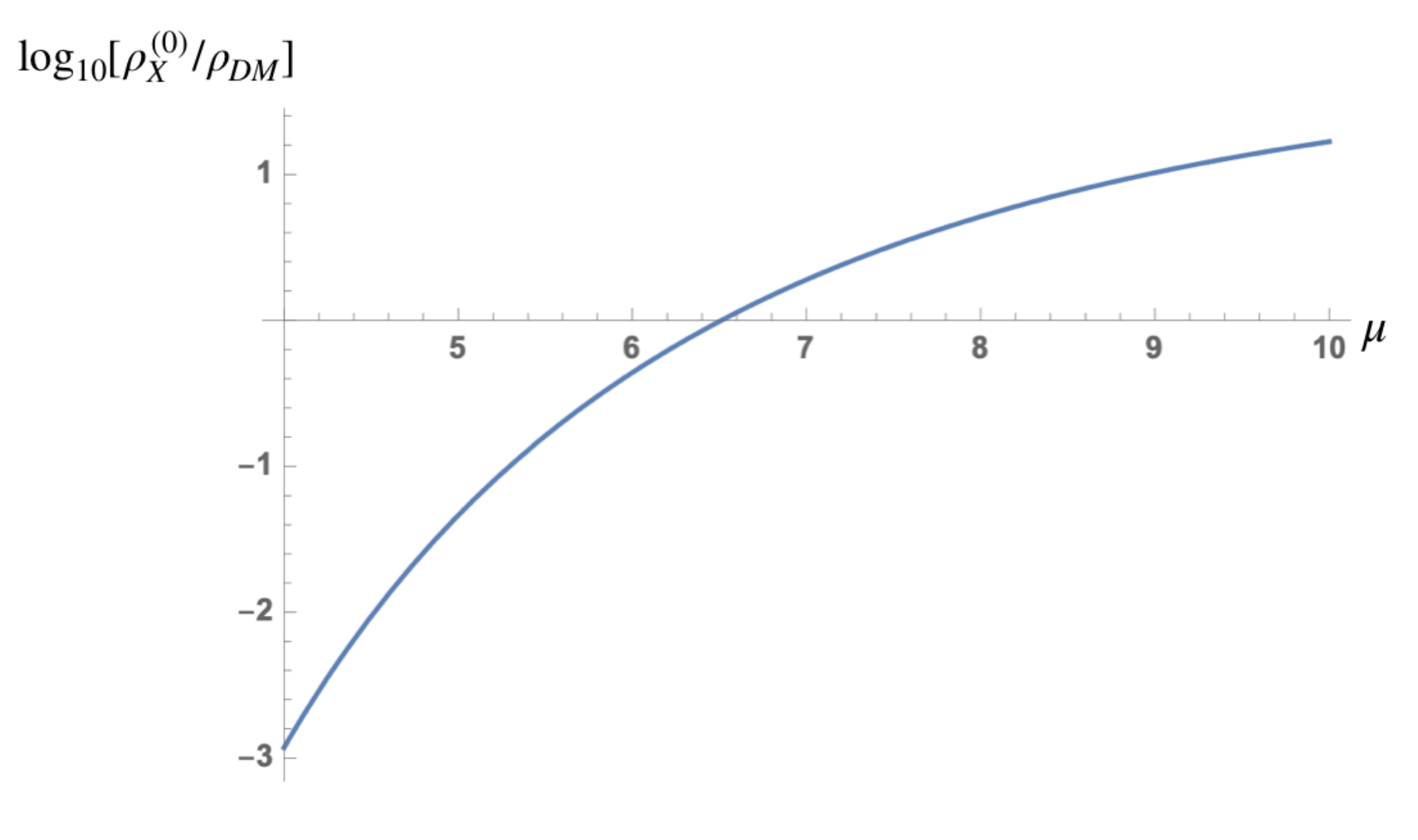}
  \caption{ Log of the ratio of the energy density of $X$-particles (\ref{rho-X-u}) to the  observed energy density of dark matter as a function 
  of $\mu =M_R/M_X$.   }
  \label{rho-of-MX}
 \end{figure}

The logarithm of the energy density of $X$-particles (\ref{rho-X-u})  with respect to the observed energy density of dark matter as a function of $M_X$
is presented in Fig.~\ref{rho-of-MX}. If $M_X \approx 5\cdot 10^{12}$ GeV, X-particles may be viable candidates for the 
carriers of the cosmological dark matter.

\section{Possible observations \label{s-obs}}
This section is outside of the scope of our paper. It contains a discussion of some rather speculative possibilities
of observation of the products of X-particle slow decay or enhanced annihilation in ultra high energy cosmic rays.
More detailed study of the phenomena considered below demand separate work and the effects may be very weak
or even non-existing. So the reader can skip this short section.

There are two possibilities to make X-particles visible: firstly, due to possible high density of  $X \bar X$-systems and, secondly, because of hypothetical instability of $X$-particles.   

According to results of this and our previous papers~\cite{Arbuzova:2018apk,Arbuzova:2018ydn}
the mass of dark matter particles, with the interaction strength typical for supersymmetric ones, can
be in the range from $10^6$ to $10^{13} $ GeV. It is tempting to find if and how they could be observed,
except for their gravitational effects on galactic and cosmological scales.

The average cosmological energy/mass density of $X$-particles in the universe is approximately  1 keV/cm$^3$,
while in galaxies it is about  1 GeV/cm$^3$. S{o their number densities should respectively be:
\begin{equation}
n_{cosm} = 10^{-12} M_6^{-1}/ {\rm cm}^3 ,\,\,\, n_{gal}  = 10^{-6} M_6^{-1}/ {\rm cm}^3 ,
\label{n-gal}
\end{equation}
where $M_6 = M_X/( 10^6 {\rm GeV} )$. 

The characteristic annihilation time in a galaxy is:
\be
\tau^{ann}_{gal} = 1/\left[\sigma_{ann} v n_{gal} \right] \approx 10^{37} M_6^3 \,
{\rm sec},
\label{tau-gal}
 \ee
where we have taken $\sigma_{ann} v \approx 10^{-2} /M_X^2$.

The total energy flux from all annihilations in the Galaxy of the size $R_{gal} \approx 10 {\rm kpc} = 10^{22} \rm{cm} $
would be
\begin{multline}
L_{gal} = n_{gal} E R_{gal} / \tau^{ann}_{gal} \approx 10^{-15} M_6^{-3}\, {\rm GeV/cm^2/sec} \\ = 
3\cdot 10^2 M_6^{-3}\, {\rm GeV/km^2/year} 
\label{L-gal}
\end{multline}
with characteristic energy of the order of $ E\sim M_X$.

The annihilation would be strongly enhanced in clusters (clumps) of dark matter~\cite{clusters}, especially
in neutralino stars~\cite{n-stars}. Based on the  latter reference, for the annihilation cross-section 
$\sigma_{ann} v = 4\cdot 10^{-42} M_6^{-2}\, \rm{cm}^2 \approx   10^{-31}M_6^{-2}\, {\rm cm}^3/{\rm  sec}$,
we can conclude that the observation of $X\bar X$-annihilation
from neutralino stars is not unrealistic. 

Due to their huge mass relic X-particles might form gravitationally bound states and then annihilate 
like positronium. Instead of fine structure constant $\alpha = 1/137$ we must use the 
gravitational coupling constant $\alpha_G = (M_X/m_{Pl})^2$. In complete analogy with para-positronium decay
the lifetime of such bound state with
respect to annihilation  would be 
\be 
\tau_G \sim (\alpha_G^5 M_X)^{-1} \approx 5\cdot  10^{23} M_{13}^{-11} \,{\rm sec},
\label{tau_G}
\ee
where $M_{13} = M_X/(10^{13} \,{\rm GeV})$.

The flux of ultra-high energy cosmic rays (UHECR) with energy $\sim 10^{21}$eV produced by the 
population of the bound states of $X\bar X$, say, from the sphere of the radius of $R=1$ Gpc would be:
\be 
F = n_{gal} R f / \tau_G = 2 \cdot10^5 {f} M_{13}^{10} \,{\rm cm^{-2} sec^{-1}},
\label{F-UhECR}
\ee
where $f$ is the fraction of bound states with respect to total number of $X$-particles.

Comparing this result with the data presented in Ref.~\cite{uhecr-DM-decay} we can conclude that 
the flux of the UHECR produced in the decay of  $X\bar X$ bound states would agree with the
data if $f \sim 10^{-11}$. 

Calculation of $f$ is subject to many uncertainties and it is not the aim of the present work. It will be done
elsewhere.

X-particle would be observable if they are unstable. Heavy $X$-particles would decay though 
formation of virtual black holes, according to  the
Zeldovich mechanism~\cite{zeld-BH-eng,zeld-BH-rus}. 
If X-particles are composite states of three fundamental constituents, as proton made of three quarks,  
their life-time with respect to virtual BH stimulated  decay  would be
\be
 \tau_{X,BH} \sim \frac{m_{Pl}^4}{ M_X^5 }\sim 10^{-13} {\rm sec} \left( \frac{10^{13}{\rm GeV} }{M_X} \right)^5 \,.
\label{tau-X-BH}
\ee
To make the time $\tau_{X,BH}$ larger than the universe age $t_U \approx 4\cdot 10^{17}$ sec we need 
$M_X <10^7$~GeV. In this case the products of the decays of X-particles with such masses could be observable 
in the flux of the cosmic rays with energy somewhat below $10^7 $ GeV.

The life-time may be further suppressed  if we apply the conjecture of Ref.~\cite{BDF} which leads to 
a strong suppression of the decay through virtual black holes for spinning or electrically charged X-particles.
However, this suppression does not operate for spinless neutral particles. Moreover it would not be efficient
enough to sufficiently suppress the decay probability of the superheavy particles of dark matter
with masses of the order of $ 10^{13} $ GeV. The decay rate may be strongly diminished if $X$-particles consist
of more than three fundamental constituents. 
For example, if X-particles
consist of six  fundamental constituents, then the decay life-time  would be
 \be
 \tau'_{X,BH} \sim \frac{m_{Pl}^{10}}{ M_X^{11} }
 \sim 10^{23} {\rm sec} \left( \frac{ 10^{13}{\rm GeV} }{M_X} \right)^5 \,.
\label{tau-prime-X-BH}
\ee
This life-time is safely above the universe age $t_U \approx 4\cdot 10^{17} $ seconds.

\section{Conclusion and discussion \label{s-concl}}

There is general agreement that the conventional Friedmann cosmology is incompatible with the existence of stable particles
having interaction strength typical for supersymmetry and heavier than several TeV. A possible way to save life of such particles,
we call them here $X$-particles,
may be a modification of the standard cosmological expansion law in such a way that the density of such heavy relics  would be
significantly reduced. A natural way to realize such reduction presents popular now Starobinsky inflationary model~\cite{star-infl}.
If the epoch of the domination of the curvature oscillations (the scalaron domination) lasted after freezing of massive species, their
density with respect to the plasma entropy could be noticeably suppressed by production of radiation from the scalaron decay.

The concrete range of the allowed mass values depends upon the dominant decay mode of the scalaron. If the scalaron is minimally
coupled to scalar particles $X_S$, the decay amplitude does not depend upon  the scalar particle mass and leads to too high energy density of 
$X$-particles, if $M_{X_S} \ll M_R$. An acceptably low density of $X_S $ can be achieved if $M_{X_S} \gtrsim M_R \approx 3\cdot 10^{13}$ GeV.

If X-scalars are conformally coupled to curvature or X-particles are fermions, then the probability of the scalaron decay 
is proportional to $M_X^2$. For sufficiently small $M_X$ the production of $X$-particles would be quite weak, so that their
cosmological energy density would be close to the observed density of dark matter if
$M_X \sim 10^6$ GeV~\cite{Arbuzova:2018ydn}.

There is another complication due to conformal anomaly, which leads to efficient decay of scalaron into massless or light gauge bosons.
There are some versions of supersymmetric theories where conformal anomaly is absent,
which were considered in Ref.~\cite{Arbuzova:2018ydn}. In the present work we have not impose this restriction and studied a model
with full strength conformal anomaly. In this case the thermalization of the cosmological plasma started from the creation of gauge bosons
and the reactions between them created all other particle species.

There are two possible processes through which $X$-particles could be produced: direct decay of the scalaron into a pair of $\bar{X} X$ 
and the  thermal production of X's in plasma. To restrict the density of $X$-particles produced by the direct decay the observed value
$M_X$ should be below $10^7$  GeV. But in this case the thermal production of X's would be too strong. We can resolve this inconsistency
 if the direct decay of the scalaron into X-particles is suppressed and due to that 
  a larger $M_X$ is allowed, so the thermal production
 would not be dangerous. The direct decay can be very strongly suppressed if X-particles are Majorana fermions, which cannot be created
 by  a scalar field in the lowest order of perturbation theory. It opens the possibility for X-particles to make proper amount of dark matter,
 if their mass is about $5 \cdot 10^{12}$ GeV. 

 Thus a supersymmetric type of dark matter particles seems to be possible if their mass is quite high from $10^6$ up to $5 \cdot 10^{12} $ GeV, or
 even higher than the scalaron mass,  $M_R = 3\cdot 10^{13}$~GeV. 
There is not chance to discover these particles in accelerator
 experiments in foreseeable future,
 but they may be observable through cosmic rays from their annihilations in high density clumps of
 dark matter, or from annihilation in their gravitationally bound two-body states, or through the products of their
 decays, since they naturally should be unstable.

\section*{Acknowledgements}
The work  was supported by the RSF Grant 19-42-02004.


\end{document}